\documentclass[a4paper]{article}

\usepackage[pages=all, color=black, position={current page.south}, placement=bottom, scale=1, opacity=1, vshift=5mm]{background}
\SetBgContents{
	
}      
\usepackage[margin=1in]{geometry} 

\usepackage{amsmath}
\usepackage{amsthm}
\usepackage{amssymb}
\usepackage{bbold}
\usepackage[utf8]{inputenc}
\usepackage{lmodern}
\usepackage[T1]{fontenc}
\DeclareUnicodeCharacter{2009}{\,}
\usepackage{hyperref}
\hypersetup{
	unicode,
	pdfauthor={Shayan Roofeh, Vahid Karimipour},
	pdftitle={The Landau-Streater Channel as a Noisy Quantum Channel},
	pdfsubject={The Landau-Streater Channel as a Noisy Channel Model},
	pdfkeywords={article, template, simple},
	pdfproducer={LaTeX},
	pdfcreator={pdflatex}
}

\usepackage[sort&compress,numbers,square]{natbib}
\theoremstyle{plain}

\theoremstyle{definition}
\newtheorem{definition}{Definition}

\usepackage{graphicx}
\graphicspath{{fig/}}
\usepackage{subcaption}
\usepackage{algorithm, algpseudocode} 
\usepackage{mathrsfs} 
\usepackage{lipsum}
\usepackage{physics}
\def\be{\begin{equation}}
	\def\ee{\end{equation}}
\def\ba{\begin{eqnarray}}
	\def\ea{\end{eqnarray}}
\def\lo{\longrightarrow}
\def\h{\hskip 1cm }

\def\la{\langle}
\def\ra{\rangle}
\def\a{\alpha}

\def\ni{\noindent}

\definecolor{darkret}{RGB}{139,0,0}
\def\tr{\textcolor{black}}
\def\tb{\textcolor{black}}
\def\bex{\begin{dinglist}{110}\dsquare}
	\def\eee{\end{dinglist}}
\def\bet{\begin{dinglist}{110}\bsquare}
	\def\bfr{\begin{mdframed}[backgroundcolor=blue!20]\vspace{0.5cm}}
		\def\efr{\vspace{0.5cm}\end{mdframed}}
	
	\title{Exact Quantum Capacity of Decohering Channels in Arbitrary Dimensions}
	\author{Shayan Roofeh$^1$ \and Vahid Karimipour$^1$}
	
	\date{
		$^1$\small{Department of Physics, Sharif University of Technology, Tehran, Iran} \\%
	}
	
\begin{document}
		\maketitle


		\begin{abstract}
			We derive exact analytical expressions for the quantum capacity of a \tr{broad subclasses of generalized dephasing channels of the form} $\Lambda(\rho)=(1-x)\rho + x D(\rho)$, where $D(\rho)$ represents a structured decoherence process. \tr{These channels are  degradable for all noise parameters and in arbitrary dimensions}, yielding closed-form, single-letter capacity formulas. Our analysis includes fully decohering, block-decohering, and weakly decohering channels, the latter involving coherence preservation within overlapping subspaces. Surprisingly, even under maximal decoherence, the channel may retain nonzero capacity due to residual coherence structure. These results provide quantitative role for  decoherence-free and partially coherent subspaces in preserving quantum information, offering guidance for encoding strategies in quantum memories and fault-tolerant quantum communication systems.
		\end{abstract}

	
\section{ Introduction}

The capacity of a quantum channel to transmit information is a central figure of merit in quantum communication and quantum information processing \cite{NielsenChuang2000,Lloyd1997,Schumacher_1996}. Yet, despite its conceptual clarity, calculating the quantum capacity remains notoriously difficult, even for seemingly simple channels. This difficulty stems from the need to optimize the coherent information over multiple uses of the channel and the lack of additivity in general \cite{Shor2004,Hastings_2009,SmithYard2008}, rendering the computation of most capacities intractable.\\

\tb{These theoretical challenges are mirrored in the experimental domain. Directly measuring a channel's quantum capacity is infeasible, as it would require exhaustive quantum process tomography followed by complex optimizations. 
Consequently, recent experimental efforts have focused on developing efficient protocols to bound capacities by estimating key channel properties, such as the average fidelity via randomized benchmarking \cite{gambetta2010randomized} or the diamond norm via direct fidelity estimation \cite{da2010direct}, moving beyond earlier demonstrations that required more extensive measurements \cite{macchiavello2004experimental, durkin2007experimental}.}\\

Exact capacity formulas are known only for a handful of special cases, such as the erasure channel \cite{Bennett_1997}, amplitude damping in limited regimes in low dimensions \cite{devetak_capacity_2005,chessa_quantum_2021, Chessa_2023}, and a few qubit Pauli channels \cite{Shirokov2006}. In higher-dimensional systems, exact results are almost nonexistent \cite{KarimipourSciRep2025,karimipour2024}. Even lower and upper bounds can be difficult \cite{devetak_capacity_2005,Wolf2007,SmithSmolin2007,Cubitt2008,Fukuda2009,FanizzaKianvashGiov2020, Roofeh_2024, RoofehKarimipour2024} to compute due to nonconvexities in the coherent information. Thus, any exact result in this domain not only deepens our theoretical understanding but also offers valuable benchmarks for real-world implementations.
In this Letter, we present a broad family of quantum channels for which the quantum capacity can be computed exactly and in closed form. These channels take the form $\Lambda(\rho)=(1-x)\rho + x D(\rho)$, where $D(\rho)$ models structured decoherence \cite{baumgratz_quantifying_2014,zurek2003,schlosshauer2005,Mani_2015,ManiRezazadehKarimipour2024}. \tr{Remarkably,  these channels are degradable in all dimensions and for all values of the noise parameter $x$}.   This includes fully decohering channels, block-decohering models that preserve coherence within subspaces, and weakly decohering channels where coherence survives across overlapping modes. Here we provide an alternative proof for their degradability and show that for all these channels, the complementary channel itself degrades the channel. We show that even under strong noise $(x = 1)$, these channels may retain nonzero quantum capacity—provided certain coherence structures are preserved \cite{PhysRevLett.84.2525,  PhysRevLett.82.2417, PhysRevLett.79.3306, Lidar_2003, Duan_1997, ManiRezazadehKarimipour2024}. This finding challenges the common intuition that strong decoherence always implies complete quantum information loss.\\

\ni We first consider the  operationally important quantum channels, of the form $$\Lambda(\rho) = (1-x)\rho + x D(\rho),$$ where $D(\rho)$ induces complete dephasing, then we study channels of the form
$$\Lambda_k(\rho) = (1-x)\rho + x D_k(\rho),$$
where now block-decoherence \cite{aberg_quantifying_2006,bischof_resource_2019,bischof_quantifying_2021,ManiRezazadehKarimipour2024,levi_quantitative_2014,Ringbauer2018,Johnston2018,xu_coherence_2020}  occurs between orthogonal subspaces of dimension $k$ and coherence is preserved in each subspace. The surprise is  when decoherence acts between subspaces rather than individual states: even under maximal noise ($x=1$), the channel retains non-zero capacity, if coherence persists within subspaces. This demonstrates that subspace-localized coherence, a form of "protected" quantum information, can defy complete erasure. Our explicit capacity formula quantifies this effect, and shows how decoherence-free subspaces \cite{kais_review_2014} contribute in the capacity of this channel. 
\tr{To go beyond these types of direct sum channels \cite{chessa_partially_2021}}, we proceed to  derive the exact capacity for a third class of quantum channels of the form
$$\Phi_k(\rho) = (1-x)\rho + x W_k(\rho),$$
where $W_k$ destroys coherence between overlapping subspaces in such a way that there is no decoherence-free subspace.  Even in this case, we see that capacity remains non-zero in the limit of strong decoherence. \\

\noindent The structure of this paper is as follows: In section \ref{Cap}, we collect some known facts about the quantum capacity of quantum information and describe why it is so difficult to obtain exact results for capacities. In  \ref{D1}, we first study the case of fully decohering channel $\Lambda$, where the dimension of each subspace is $1$. After proving its degradability property we proceed to calculate its quantum capacity. We then go on in section \ref{Dk} to consider the case where decoherence occurs between orthogonal subspaces of dimension $k$. The proofs and methods of calculation are similar to the case of $k=1$ with slight differences. Finally in section \ref{kdeco}, we study the case of decoherence among overlapping subspaces. The basic results are shown in figures (\ref{Capacities}) and (\ref{fig:coherent-info}). The paper is concluded with a discussion.

\section{Preliminaries on quantum capacities}
\label{Cap}
\ni A central concept in the study of quantum channels is the quantum capacity, which quantifies the maximum rate at which quantum information can be reliably transmitted through a channel. The quantum capacity is a key measure of channel performance  \cite{barnum_information_1998, lloyd_capacity_1997,devetak_private_2005}, directly impacting the efficiency and reliability of quantum communication systems. However, calculating the quantum capacity is notoriously difficult, even for relatively simple channels like the depolarizing channel. This difficulty arises from the generic super-additivity property of most quantum channels.  This is one of the most challenging aspects of quantum capacity. \tr{Unlike classical capacities of classical channels, where the capacity of two independent channels is simply the sum of their individual capacities, the quantum or even the classical capacity of the tensor product of two quantum channels  can be greater than the sum of the capacities of the individual channels \cite{Hastings_2009}. } This  behavior significantly complicates the calculation of quantum capacity because it requires considering not just single uses of the channel, but also the possibility of using entangled input states across multiple channel uses. As a result, the  quantum capacity of a channel 
can only be determined by optimizing the coherent information over an infinite number of uses of the channel \cite{cubitt_2015}, a task which in general is intractable.  More precisely the quantum capacity of a channel ${\cal E}$ should be calculated via \cite{devetak_private_2005}:
\begin{equation}
\label{Qcapacitydef}
	\begin{split}
		Q({\cal E}) =
		\lim_{n \to \infty}  \; \frac{1}{n} Q^1(
		{\cal E}^{\otimes n}),
	\end{split}     
\end{equation}
\tb{where $n$ is the number of uses of the channel, $Q^1({\cal E})=\max_{\rho} I_c({\cal E},\rho)$ and the coherent information is given by $I_c({\cal E},\rho):=S({\cal E}(\rho))-S({\cal E}^c(\rho))$, where ${\cal E}^c$ is the complementary channel.} Here $S(\rho)$ is the von-Neumann entropy $S(\rho):=-\tr \rho \log \rho$. In general  the need for the limiting procedure over an infinite number of channel uses, i.e. the so-called regularization, renders an exact calculation of the quantum capacity extremely difficult (or virtually impossible). For this reason,  although there are certain general theorems on the capacity of quantum channels \cite{Hastings_2009,fern_lower_2008,divincenzo_quantum-channel_1998,leditzky_quantum_2018,bausch_error_2021,siddhu_positivity_2021,smith_additive_2007}, \tr{there are  few channels for which the quantum capacity is known in closed form, i.e. bit-flip and phase flip channels and quantum erasure channel \cite{Schumacher_1996, devetak_private_2005, Bennett_1997}, for the full range of their parameters, see also \cite{chessa_quantum_2021,Chessa_2023}, where the capacity of multi-level amplitude damping channel has  been calculated in closed form in a certain range of parameters. It is also known that the 
amplitude damping channels for qubits $A_\gamma$, where $A_\gamma(\rho)= A_0\rho A_0^\dagger + A_1\rho A_1^\dagger$ \footnote{
$A_0=\begin{pmatrix}1&0\\ 0 &\sqrt{1-\gamma}\end{pmatrix}$ and $A_1=\begin{pmatrix}0&\sqrt{\gamma}\\ 0 &0\end{pmatrix}$ .}
 is   degradable  for $\gamma\in [0,\frac{1}{2}] $ and anti-degradable for $\gamma\in [\frac{1}{2},1]$. In passing we stress that being degradable or anti-degradable is not a necessary condition for exact calculability of the quantum channels and indeed there  are  channels which lack these properties and yet  their quantum capacity can be determined exactly. For examples of these chanels see \cite{bhalerao2023improving}}. \\

\ni We should also note that for  almost any other quantum channel, one suffices to obtain lower and upper bounds for the quantum capacity \cite{kianvash1, kianvash2, poshtvan_capacities_2022, fern_lower_2008, Chessa_2019}
. Even these bounds are not easy to obtain due to the absence of convexity properties in the coherent information.  \\

\ni An important development in this direction was made by introducing the concepts of degradability of quantum channels \cite{devetak_capacity_2005,cubitt_structure_2008}.  

\begin{definition} \cite{devetak_capacity_2005,cubitt_structure_2008} \label{definitonofdeg} Consider a quantum channel ${\cal E}: A
	\to B$ and its complementary channel ${\cal E}^c: A\to E$, where $A$ and $B$ denote the domain and target systems and $E$ the auxiliary system playing the role of environment. The channel ${\cal E}$ is said to be degradable if there exists a quantum channel $\mathcal{M}: B \to E$ such that $\mathcal{M} \circ {\cal E} = {\cal E}^c$. Conversely it is called anti-degradable if  there is a quantum channel $\mathcal{N}: E\to B$ such that $\mathcal{N} \circ {\cal E}^c = {\cal E}$.\end{definition}

\noindent For a  degradable channel the additivity property holds and we have  $Q({\cal E})=Q^1({\cal E})$ \cite{devetak_capacity_2005}, and in this case the calculation of the quantum capacity becomes a convex optimization problem. The capacity of an anti-degradable channel is zero. \\ 

\ni What we will show in this paper is to prove that the decohering channels $\Lambda, \Lambda_k$ and $\Phi_k$ are all degradable for all values of $x$ in every dimension and use this property to find their capacities in closed form. The results are shown in figures (\ref{Capacities}) and (\ref{fig:coherent-info}). 

\section{The fully-decohering channel $\Lambda$}
\label{D1}
\begin{figure}[H]
	\centering
	\includegraphics[width=14cm]{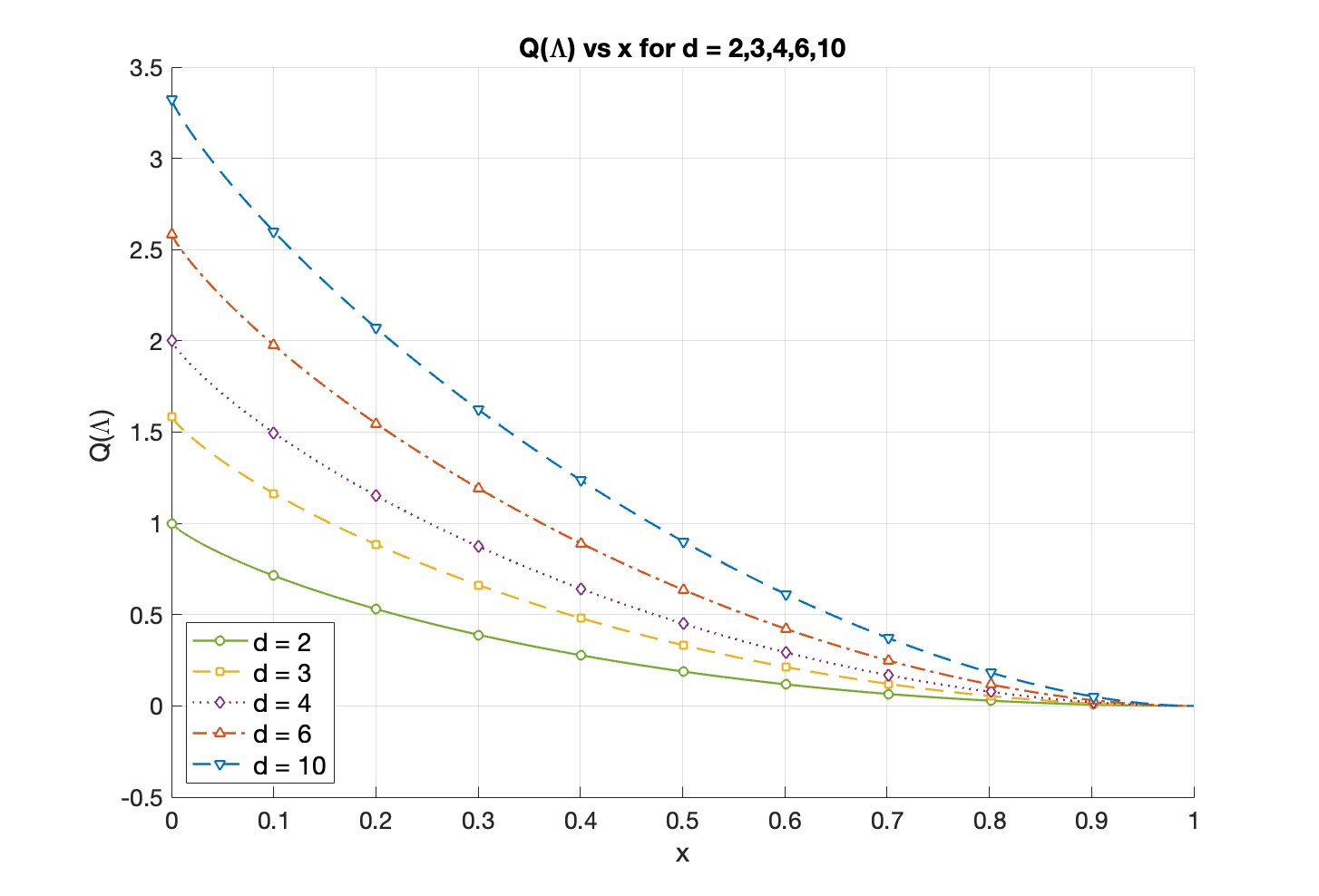}\vspace{0.0cm}
	\caption{\tb{The quantum capacities of the fully decohering  $\Lambda$ (eq. \ref{decorenceQcapa}) for dimensions $d\in\{2,3,4,6,10\}$ as a function of the noise parameter $x$.   Naturally in all dimensions, when complete decoherence occurs, the capacity vanishes. }  }
	\label{eq21}
\end{figure}

\tb{We consider the $d-$ dimensional channel $\Lambda$ (i.e. acting on density matrices defined on a $d$ dimensional Hilbert space $H_d$)  }
\be
\Lambda(\rho)=(1-x)\rho + x\ D(\rho)
\ee
where $x\in [0,1]$ and $D(\rho)=\sum_{i=1}^d \rho_{ii}|i\ra\la i|$ shows the fully decohered state of $\rho$. 
This channel shows the effect of decohering noise which happens with probability $x$. \tr{The channel $D$ is  a special case of the generalized dephasing channel ${\cal D}_g $ \cite{khatri2023positivity} defined as follows:
\be
D_g(\rho)=\sum_{i,j=1}^d \rho_{ij}\la \psi_i|\psi_j\ra |i\ra\la j|,
\ee
where $|\psi_i\ra$ are arbitrary states. In our case, we have taken them to be orthonormal. }
The first basic property of this channel, which will play a central role in proving the degradability of the channel  is that for all input states $\rho$,
$
D(\Lambda(\rho))=D(\rho).
\label{property1}
$
or simply
\be\label{one}
D\circ \Lambda=D.
\ee
The Kraus operators of the channel are given by
\be
A_0=\sqrt{1-x}\ I\h A_i=\sqrt{x}\ |i\rangle\langle i|.
\label{decKraus}
\ee
From the isometric extension, we find that the complementary channel of $\Lambda$ is given by 
\be
\Lambda^c(\rho)=\sum_{i,j}Tr(A_i \rho A_j^\dagger)\ |i\rangle\langle j|.
\label{decComp}
\ee
Therefore we find from (\ref{decKraus}) and (\ref{decComp}), 
\be
\Lambda^c(\rho)=\begin{pmatrix} (1-x)Tr(\rho)& a\rho_{11}&a\rho_{22}&\cdot &a\rho_{dd}\\ a\rho_{11}& x\rho_{11}&0&0 & 0\\
	a \rho_{22}& 0& x\rho_{22}&0 & 0\\
	\cdot&0&0&\cdot&0&\\
	a \rho_{dd}&0&0&0&x \rho_{dd},
\end{pmatrix}
\ee
where $a=\sqrt{x(1-x)}$.
The second crucial property is that $\Lambda^c(Y)$ depends only on the diagonal entries of $Y$, that is, for any input state $\rho$ 
\be
\label{property2}
\Lambda^c=\Lambda^c\circ D.
\ee
 The combination of (\ref{one}) and (\ref{property2})leads to the following chain of identities
 \be
\Lambda^c\circ \Lambda=(\Lambda^c\circ D)\circ \Lambda= \Lambda^c\circ (D\circ \Lambda)= \Lambda^c\circ D=\Lambda^c,
\ee
where in the first equality we have used (\ref{property2}), in the second, we have used (\ref{one}) and then in the third equality, 
we have again used (\ref{property2}). \tr{Thus we see that  the channel $\Lambda$ is degradable and $\Lambda^c$ in fact degrades $\Lambda$, that is
$$\Lambda^c\circ \Lambda=\Lambda^c.$$ 
The above equation was first proved in \cite{khatri2023positivity} for the generalized dephasing channel of which our channel $D$ is a special case. }
 Note that this is entirely different from self-complementary channels \cite{smaczynski2016selfcomplementary}, i.e. those for which $\Lambda^c=\Lambda$. These channels are both degradable and anti-degradable and hence have zero capacity. The degradability of the decohering channel $\Lambda$, has now important consequences, namely that its full quantum capacity is the same as its one-shot capacity $Q^{(1)}(\Lambda).$
 Moreover, degradability also implies that the coherent information $J(\Lambda,\rho)$ is a concave function of $\rho$. These two results  greatly facilitate the exceedingly difficult task of optimization in eq. (\ref{Qcapacitydef}). First we can bypass the almost impossible task of regularization over infinitely many uses of the channel. Second we use concavity of the coherent information to do the optimization in a direct analytical way. \tb{To do this, we note that due to the Schur-Horn theorem in matrix theory \cite{Schur1923, Horn1954},} the diagonal entries of any Hermitian matrix are majorized by its eigenvalues. Thus we have 
\be\label{dD}
d(A)\prec \lambda(A).
\ee
\tb{We remind the reader that a probability distribution $\{p\}=\{p_1\geq p_2\geq \cdots p_d\}$ is  majorized by another probability distribution $\{q\}=\{q_1\geq g_2\geq \cdots q_d\}$, if the following relations hold:
\ba
p_1&\leq& q_1\cr
p_1+p_2&\leq& q_1+q_2\cr
p_1+p_2+p_3&\leq& q_1+q_2+q_3\cr
\cdots &\leq &\cdots.
\ea
It is a well-known property of the Shannon entropy that in this case, $H(q)\leq H(p)$, where$H(p)=-\sum_i p_i\log_2 p_i$. Intuitively this means spreading the probability distribution, increases the Shannon entropy. }Therefore from (\ref{dD}), we find that  leads to 
\be
S(\Lambda(\rho))\leq S(D(\Lambda(\rho))).
\ee
For the complementary channel we also know from $\Lambda^c\circ D=\Lambda^c$ that 
\be
S(\Lambda^c(\rho))=S(\ \Lambda^c(D(\rho))\ ).
\ee
These two results together imply that to maximize the coherent information, it is sufficient to only search the space of diagonal matrices $\rho_d$, namely 
\be
I_c(\Lambda, \rho)\leq I_c(\Lambda, \rho_d).
\ee
We now note that the convexity property of a function does not always imply that the maximum is attained at the maximally mixed state. A simple counter-example is the function $f(\rho)=\tr(X\rho)$ for a Hermitian operator $X$, whose maximum is attained at $\rho=|\lambda_{0}\ra\la \lambda_{0}|$, where $|\lambda_{0}\ra$ is the eigenvector corresponding to the largest eigenvalue of $X$.  To reach this conclusion for the coherent information $I_c(\Lambda,\rho)$, we invoke  the covariance property of the channel, namely 

\be
\Lambda(X\rho X^\dagger)=X\Lambda(\rho)X^\dagger,
\ee
where $\rho$ is any matrix (diagonal or not) and $X$ is the shift operator
\be
X=\sum_i |i+1\rangle\langle i|.
\ee
\tr{Like any other channel \cite{poshtvan_capacities_2022}, the covariance of the channel imposes the covariance of the complementary channel \cite{holevo2005complementarychannelsadditivityproblem} which in the present case takes the form }
\be
\Lambda^c(X\rho X^\dagger)=U_x\Lambda^c(\rho)U_x^\dagger,
\ee
where $U_x=|0\ra\la 0|\oplus X$. This means that $I_c(\Lambda,\rho)=I_c(\Lambda,X\rho X^\dagger)$ for any density matrix. 
We now note that for  diagonal density matrix $\rho_d$, it holds that
\be
\frac{1}{d}\sum_{i=1}^d X^i \rho_d {X^\dagger}^i=\frac{1}{d}I_d.
	\ee
Using this covariance and the concavity of the coherent information we can now write 
\be
I_c(\Lambda,\rho_d)=\frac{1}{d}\sum_{i=1}^d I_c(\Lambda,\rho_d)=\frac{1}{d}\sum_{i=1}^d I_c(\Lambda, X^i\rho_d {X^\dagger}^i)\leq I_c(\Lambda,\frac{1}{d}\sum_{i=1}^d X^i\rho_d {X^i}^\dagger) =I_c(\Lambda,\frac{1}{d}I_d)
\ee
This means that the maximum is attained on the maximally mixed state and hence, due to unitality of the channel $\Lambda(\frac{I}{d})=\frac{I}{d}$, we have
\be
Q(\Lambda)=S(\frac{I}{d})-S(\Lambda^c(\frac{I}{d})).
\ee
It is readily found that 
\be
\Lambda^c(\frac{I}{d})=(1-x)|0\ra\la 0|+\frac{\sqrt{x(1-x)}}{d}\sum_{i=1}^d (|i\ra\la 0|+|0\ra\la i|)+ \frac{x}{d}\sum_{i=1}^d |i\ra\la i|,
\ee
or in explicit matrix form
\be
\Lambda^c(\frac{I}{d})=\begin{pmatrix}(1-x)& \xi\la v_0|\\ \xi| v_0\ra& \frac{x}{d}I_d\end{pmatrix}
\ee
where $\xi=\sqrt{\frac{x(1-x)}{d}}$, $|v_0\ra=\frac{1}{\sqrt{d}}\sum_{i=1}^d|i\ra$ is the maximally coherent state and $I_d$ is the $d-$ dimensional identity matrix. The eigenvalues of this matrix are  $$\{0, \frac{d+x-dx}{d},\frac{x}{d}\} $$ where the last eigenvalue has degeneracy of $d-2$, (i.e. a total of $d$ eigenvalues). Collecting all terms, we find the quantum capacity of the channel $\Lambda_x$ in closed form as
\be
Q(\Lambda)=\log d + \frac{d-1}{d}x\log \frac{x}{d}+(1-x+\frac{x}{d})\log(1-x+\frac{x}{d}),
\label{decorenceQcapa}
\ee
\tb{where in accordance with the usual notation in  information theory, all the logarithms are at base 2. }
This is the exact capacity of the channel $\Lambda$ which smoothly drops from $\log_2 d$ for the identity channel to $0$ for the fully decohering channel, as expected. When $x=1$, the channel output is a classical mixture and has no superposition. Figure (\ref{eq21}) shows a plot of these capacities for various dimensions as a function of $x$.\\

\section{The Block-decohering channel $\Lambda_k$}
\label{Dk}
\begin{figure}[H]
	\centering
	\includegraphics[width=14cm]{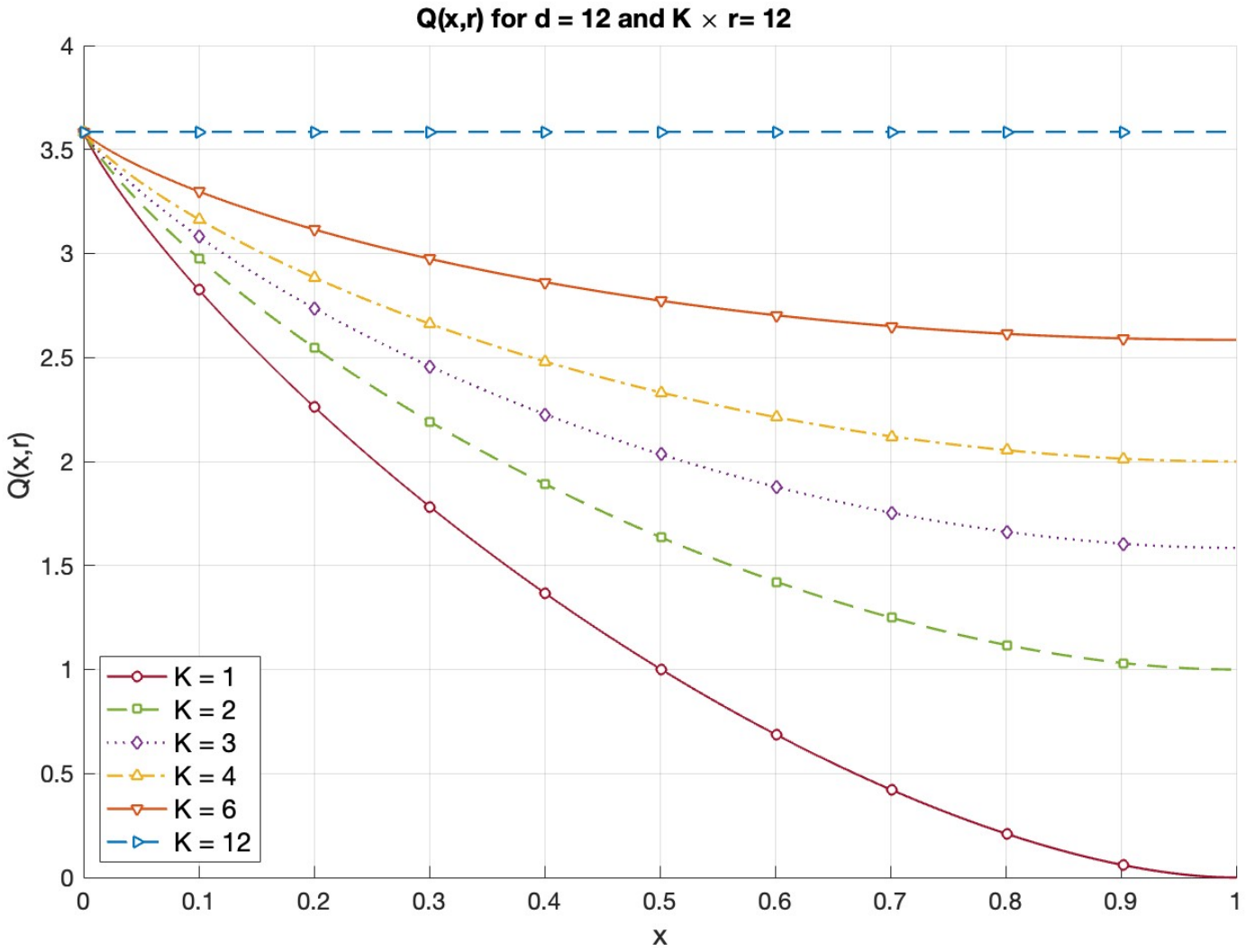}\vspace{-1cm}
	\caption{\tb{The quantum capacities of the block-decohering channels $\Lambda_k$ (eq. \ref{e33}) for $d=12$ for various values of $k$. For each $k$, the number of blocks is given by $r=\frac{12}{k}$. It is seen that even when $x=1$, full decoherence among subspaces, allows a nonzero quantum capacity. This is the capacity of quantum information transfer, when a suitable encoding into decoherence-free subspaces is used.  } }
	\label{Capacities}
\end{figure}

\tr{We now extend this method to the channel in which decoherence occurs between subspaces of dimension $k$ and coherence in each subspace is maintained. These are instances of Partially Coherent Direct Sum channels \cite{chessa_partially_2021} } and are defined  by the map
\be
\Lambda_k(\rho)=(1-x)\rho + xD_k(\rho).
\ee
Here $D_k$ depicts the decoherence map between blocks

\be
D_k(\rho)=\begin{pmatrix} \rho_{1}& \cdot &\cdot&\cdot&\cdot\\ \cdot& \rho_{2}&\cdot&\cdot&\cdot\\
	\cdot& \cdot&\rho_{3} & \cdot&\cdot\\\cdot& \cdot&\ & \cdot&\cdot\\
	\cdot&\cdot&\cdot&\cdot&\rho_{r}
\end{pmatrix}
\ee
where each $\rho_i$ is the projector of the density matrix onto the $i-th$ subspace of dimension $k$. The dimension of the full Hilbert space is taken to be $d=kr$, where $k$ is the dimension of each subspace and $r$ is the number of orthogonal subspaces, i.e. number of blocks. As a map $D_k(\rho)$ is given by 
\be
D_k(\rho)=\sum_{i=1}^r \pi_i\rho \pi_{i},
\ee
where $\pi_1=|1\ra\la 1|+\cdots |k\ra\la k| $,  $\pi_2=|k+1\ra\la k+1|+\cdots |2k\ra\la 2k|$ and so on. 
The complementary channel now becomes 
\be
\Lambda_k^c(\rho)=\begin{pmatrix} (1-x)Tr(\rho)& aTr(\rho_1)&aTr(\rho_2)&\cdot &aTr(\rho_r)\\ aTr(\rho_1)& xTr(\rho_{1})&0&0 & 0\\
	aTr(\rho_2)& 0& xTr(\rho_{2})&0 & 0\\
	\cdot&0&0&\cdot&0&\\
	a Tr(\rho_r)&0&0&0&x Tr(\rho_{r}),
\end{pmatrix}
\ee
where $a=\sqrt{x(1-x)}$. 
\tr{Both of the basic properties (\ref{property1}) and (\ref{property2}) hold also for this channel from which we find that this channel is also degradable. This results are in accord with the ones reported in \cite{chessa_partially_2021}. }Moreover, the Schur-Horn theorem is used to show that optimization can be performed by searching only among the diagonal density matrices $\rho_d$. 
The channel now has an extra symmetry, that is it is covariance under two types of unitary operations, that is: 
\be
\Lambda_k(\mathbb{X}\rho\mathbb{X})=\mathbb{X}\Lambda_k(\rho)\mathbb{X}, \h \Lambda_k(\Omega\rho\Omega)=\Omega\Lambda_k(\rho)\Omega, 
\ee
where 
\be
\mathbb{ X}=\begin{pmatrix} X_k& \cdot &\cdot&\cdot&\cdot \\ \cdot  & X_k&\cdot&\cdot&\cdot\\
	\cdot& \cdot&X_k & \cdot&\cdot\\ \cdot& \cdot&\cdot & \cdot&\cdot\\
	\cdot&\cdot&\cdot&\cdot&X_k
\end{pmatrix}
\ee
which affects shift in each subspace and 
\be
\Omega=\begin{pmatrix} \cdot& \cdot &\cdot&\cdot&I_k\\ \ I_k & \cdot&\cdot&\cdot&\cdot\\
	\cdot& I_k&\cdot & \cdot&\cdot\\ \cdot& \cdot&\cdot & \cdot&\cdot\\
	\cdot&\cdot&\cdot&I_k&\cdot
\end{pmatrix}
\ee
which affects shifts among subspaces. We now use this symmetry to show that for an arbitrary diagonal matrix $\rho_d$, 
\be
\frac{1}{d}\sum_{i=1}^k\sum_{j=1}^r U_{ij}\rho_d U_{ij}^\dagger = \frac{I}{d},
\ee
where $U_{ij}=\mathbb{X}^i\  \Omega^j$.
In the same way as for the channel $\Lambda_1$, concavity of the coherent information now yields that 
\be
\max_{\rho} I_c(\Lambda_k,\rho)=S(\Lambda_k(\frac{I}{d}))-S(\Lambda_k^c(\frac{I}{d}))
\ee
where $d=rk$ is the dimension and $\Lambda_k^c$ has exactly the same form as (\ref{decorenceQcapa}), with $d$ replaced by $r$. The final result is given by 
\be\label{e33}
Q(\Lambda_k)=\log k+\log r + \frac{r-1}{r}x\log \frac{x}{r}+(1-x+\frac{x}{r})\log(1-x+\frac{x}{r}).
\ee
The capacity of the channels at $x=0$ is given by $Q(\Lambda(x=0))=\log_2d$ and drops to $Q(\Lambda(x=1))=\log_2k$ . This is a reflection of the fact that we can encode the input states into those subspaces which are not decohered. 
Plots of capacities of these channels is shown in figure (\ref{Capacities}) for the case where dimension $d=12$ and the number of blocks increase from $k=1$ (complete coherence) to $k=12$ (no coherence). 

\section{The Weakly-Decohering Channel $\Phi_k$}\label{kdeco}
The non-zero capacity of block-decohering channels in the previous section may be attributed to the presence of decoherence-free subspaces, namely the $k-$ dimensional decoherence-free subspaces which are not affected by the decohering noise. As we will see in this section, in the absence of these decoherence-free subspaces, and provided that some degree of coherence exists, the channel can still have non-zero quantum capacity. In this way, we are considering a noise which is not so strong to  annihilate all the off-diagonal elements. Rather, we assume an equal number of minor diagonals above and below the main diagonal are also kept intact while all the other off-diagonal elements vanish. \tb{As an example, for $d=6$, an example of such weak decoherence \cite{Ringbauer2018, Johnston2018, Sperling2015, Killoran2016}}  is the following: 

\[\rho\lo W_2(\rho):=
\begin{pmatrix}
	\rho_{11} & \rho_{12} & 0         & 0         & 0         & \rho_{16}       \\
	\rho_{21} & \rho_{22} & \rho_{23} & 0         & 0         & 0         \\
	0         & \rho_{32} & \rho_{33} & \rho_{34} & 0         & 0         \\
	0         & 0         & \rho_{43} & \rho_{44} & \rho_{45} & 0         \\
	0         & 0         & 0         & \rho_{54} & \rho_{55} & \rho_{56} \\
	\rho_{61}         & 0         & 0         & 0         & \rho_{65} & \rho_{66}
\end{pmatrix}.
\]
Note that for simplicity of calculation we assume a periodic boundary condition on the states, i.e.  $|d+1\ra\equiv |1\ra$, that is why in this example the two elements in the corner are non-vanishing. 
There is a related concept of coherence, called $k-$ coherence \cite{johnston_evaluating_2018,ManiRezazadehKarimipour2024}, which is somewhat more general than this, but we will not consider it here. The original notion of incoherence \cite{senitzky_incoherence_1962}, which  
defines incoherent states as diagonal density matrices in a specific basis, has been generalized to multi-level or k-coherence \cite{ManiRezazadehKarimipour2024}. In this generalized setting, a state $|\psi\ra=\sum_{i=1}^d \psi_i|i\ra$
is said to have coherence at level k, if exactly $k$ of the coefficients $\psi_i$ are non-zero.
Thus an incoherent state has coherence at level $1$, and a state like $|\psi\ra=a|0\ra+b|1\ra$ in a $d-$ dimensional Hilbert space  has coherence at level $2$ and so on. The generalization to mixed states is done by defining the states with coherence level $k$ to be the convex combination of all pure states whose coherence level is less than or equal to $k$. The notion of weak-coherence that we adopt here is different in that it assigns an order to the basis states. For example in a $d-$ dimensional space, if for some physical reason, say the photon number in each state, we order the states as $|0\ra, |1\ra, |2\ra, \cdots$, two states of the form $|0\ra+|1\ra$ and $|1\ra+|2\ra$ are equally affected by decoherence, but a state like $|1\ra+|3\ra$ completely loses its coherence. \\

\ni  In this section, we consider decohering channels which with a certain probability decohere the input state to level-$k$ coherent states of the type we defined above. That is, we consider channels of the form 
\be
\Phi_k(\rho)=(1-x)\rho+ x W_k(\rho),
\ee
where $W_k(\rho)$ is a weakly coherent density matrix at level $k$. \tr{This can be seen an instance of a generalized dephasing channel \cite{khatri2023positivity} with overlaps among auxiliary states that can take the values $1-x$ or $0$.}
Physically this means that the channel mitigates all coherences among the basis states which are further apart than $k$ in an ordered set of indices $\{|1\ra,|2\ra,\cdots |d\ra\}$. The Kraus operators of this channel are:
\be
\label{GenralKraus}
A_0=\sqrt{1-x}I,\h A_i=\sqrt{\frac{x}{k}}\sum_{l=0}^{k-1}|i+l\ra\la i+l|.
\ee
In other words, each $A_i$ is proportional to the projection operator on a $k-$ dimensional subspace, but this time, in contrast to the previous section, the subspaces are not orthogonal and they overlap on smaller subspaces.  We first note that the basic property (\ref{property1}) still holds here, that is 
\be
D\circ \Phi_k=\Phi_k\circ D=D.
\ee
For the second property (\ref{property2}), we show that for the channel $\Phi_k$, the complementary channel $\Phi_k^c $ still depends on the diagonal elements of $\rho$. To see this, it is sufficient to note that the Kraus operators $I$, $A_i$ and the products $A_iA_j$ are all projectors on subspaces and hence  extract the diagonal elements of $\rho$, when inserted in equation (\ref{decComp}) for deriving the complementary channel. Therefore we have
\be
\Phi_k^c=\Phi_k^c\circ D.
\ee
Combining these two equations, as before, we find
\begin{equation}
	\Phi_k^c\circ \Phi_k=(\Phi_k^c\circ D)\circ \Phi_k =
	\Phi_k^c\circ (D\circ \Phi_k) =\Phi_k^c\circ D=\Phi_k^c 
\end{equation}
\tr{which shows that the channel $\Phi_k$ is also degradable, in accordance with the degradability of generalized dephasing channel \cite{khatri2023positivity}. } Invoking Schur-Horn theorem as before, we need only search among the diagonal density matrices. We also note from (\ref{GenralKraus}) that $XA_0X^\dagger=A_0,$ and $XA_iX^\dagger = A_{i+1}$, which show that the channel is covariant under shift operator $X$, that is 
\be
\Phi_k(X\rho X^\dagger)=X\Phi_k(\rho)X^\dagger.
\ee
Combining these, we find
from the concave property of the coherent information that the maximum value is obtained at $\rho_d=\frac{I}{d}$. Thus we have 
$$ Q(\Phi_k)=\max_{\rho_d}=I_c(\Phi_k, \frac{I}{d}).$$ 
To proceed we define 
$${\cal M}:=\Phi_k(\frac{I}{d})$$ and calculate the following elements:
\ba\label{m00}
{\cal M}_{0,0}&=&\frac{1}{d}\tr(A_0A_0)=\frac{1-x}{d}\cr
{\cal M}_{0,i}&=&\frac{1}{d}\tr(A_0A_i)=\frac{1}{d}\sqrt{kx(1-x)}.
\ea
To calculate the other elements, we first note in view of the fact that the operators $A_i$ are respectively proportional to projectors of the form given in (\ref{GenralKraus}), $\tr(A_iA_{j})$ depends only $|j-i|$ and not on individual $i$ and $j$ separately. Thus we write
\be\label{mij}
{\cal M}_{i,i+s}=\frac{1}{d}\tr(A_iA_{i+s})=:xb_s
\ee
where due to the periodic  condition of indices $s\equiv s+d$, we find after a close examination that  $b_s$ is given by
\be
\label{functionofB}
b(s):=\begin{cases}
	k-s& \text{if }\ \  0\leq s\leq k-1, \\
	0	& \text{if }\ \  k\leq s\leq d-k,\\
	s+k-d&\text{if} \ \ d-k+1\leq s\leq d-1.
\end{cases}
\ee

\ni\tb{ Putting everything together from equations (\ref{m00},\ref{mij}) and \ref{functionofB}),} we find that 
\begin{equation}
\label{complementarygeneral}
{\cal M}=\left[
\begin{array}{c|c}
	1-x & \xi\sqrt{k} \bra{v_0} \\ \hline
	\xi\sqrt{k} \ket{v_0} & x B
\end{array}
\right]
\end{equation}
where again $\xi= \sqrt{\frac{x(1-x)}{d}}$ and $\ket{v_0}=\frac{1}{\sqrt{d}}\sum_{i=1}^d \ket{i}$. and the matrix $B$ is a circulant matrix given by
\be
B=
	\mathrm{Circ}(b_0, b_1, \dots, b_{d-1}) =
	\begin{bmatrix}
		b_0     & b_1     & b_2     & \cdots & b_{d-2} & b_{d-1} \\\\
		b_{d-1} & b_0     & b_1     & \cdots & b_{d-3} & b_{d-2} \\\\
		b_{d-2} & b_{d-1} & b_0     & \cdots & b_{d-4} & b_{d-3} \\\\
		\vdots  & \vdots  & \vdots  & \ddots & \vdots  & \vdots  \\\\
		b_1     & b_2     & b_3     & \cdots & b_{d-1} & b_0
	\end{bmatrix}.
\end{equation}
The matrix $B$, being circulant, has the structure 
$$B=\sum_{s=0}^{d-1}b_sX^s,$$
where $X$ is the shift operator. The eigenvalues of $X$ are roots of unity in the form $\omega^m$ corresponding to the eigenvector $|v_m\ra=\frac{1}{\sqrt{d}}\sum_{i=1}^d \omega^{-im}|i\ra$,  ($m=0,1,\cdots d-1$), where $\omega=e^{\frac{2\pi i }{d}}$.   Thus the eigenvalues of $B$ are of the form
\be
\lambda_m=\sum_{s=0}^{d-1}b_s \omega^{ms},\h m=0,1,2,\cdots d-1.
\ee
In view of (\ref{functionofB}), this turns out to be 
\be
\lambda_m =  \frac{1}{dk}\left[\frac{ \sin{\left(\frac{k\pi m}{d}\right)}}{ \sin{\left(\frac{\pi m}{d}\right)}}\right]^2.
\label{Lambda}
\ee
To find the quantum capacity of $\Phi_k$, we now need to diagonalize the matrix  ${\cal M}\equiv \Phi_k^c(\frac{I}{d})$, as given in (\ref{complementarygeneral}). 
We note that $|v_0\ra$ is an eigenvector of $X$ and hence of $B$ with eigenvalue equal to $\lambda_0=\frac{k}{d}$. To find the spectrum of this matrix, we note that for any eigenvector $|v_{m\ne 0}\ra$,  the vector
\begin{equation}
	\begin{bmatrix}
		0\\
		|v_m\ra
	\end{bmatrix}
\end{equation}
is an eigenvector of ${\cal M}$ with eigenvalue equal to $x\lambda_m$. 
The other two eigenvectors lie in the span of $\ket{v_0}$ and $\ket{0}$:
\begin{equation}
	{\cal M}(\a \ket{0}+\ket{v_0})=(\a (1-x)+d \zeta) \ket{0}+ (\a\zeta + \frac{k}{d} x) \ket{v_0}
\end{equation}
The corresponding eigenvalue equation becomes, the solution of which yields two eigenvalues, namely 
\[
\{0,\ \  \frac{kx}{d}+(1-x)\}.
\]
As a result, the set of eigenvalues of ${\cal M}=\Phi_k^c(\frac{I}{d})$ is:
\[
\{x \lambda_1, x \lambda_2, \ldots, x\lambda_{d-1}\} \cup \left\{0,\frac{kx}{d}+(1-x)\right\}
\]
where $\lambda_m$ is given in (\ref{Lambda}). This leads to the final expression of the quantum capacity, namely  
We can now evaluate the quantum capacity of $\Phi$:
\be\label{e40}
Q(\Phi_k)= I_c\left(\Phi_k, \frac{I}{d}\right)= \log{d}+ \sum_{i=1}^{d-1} x\lambda_i \log (x\lambda_i)+\left[\frac{kx}{d}+(1-x)\right] \log{\left[\frac{kx}{d}+(1-x)\right]}
\ee

\begin{figure}[H]
	\centering
	\includegraphics[width=0.8\textwidth]{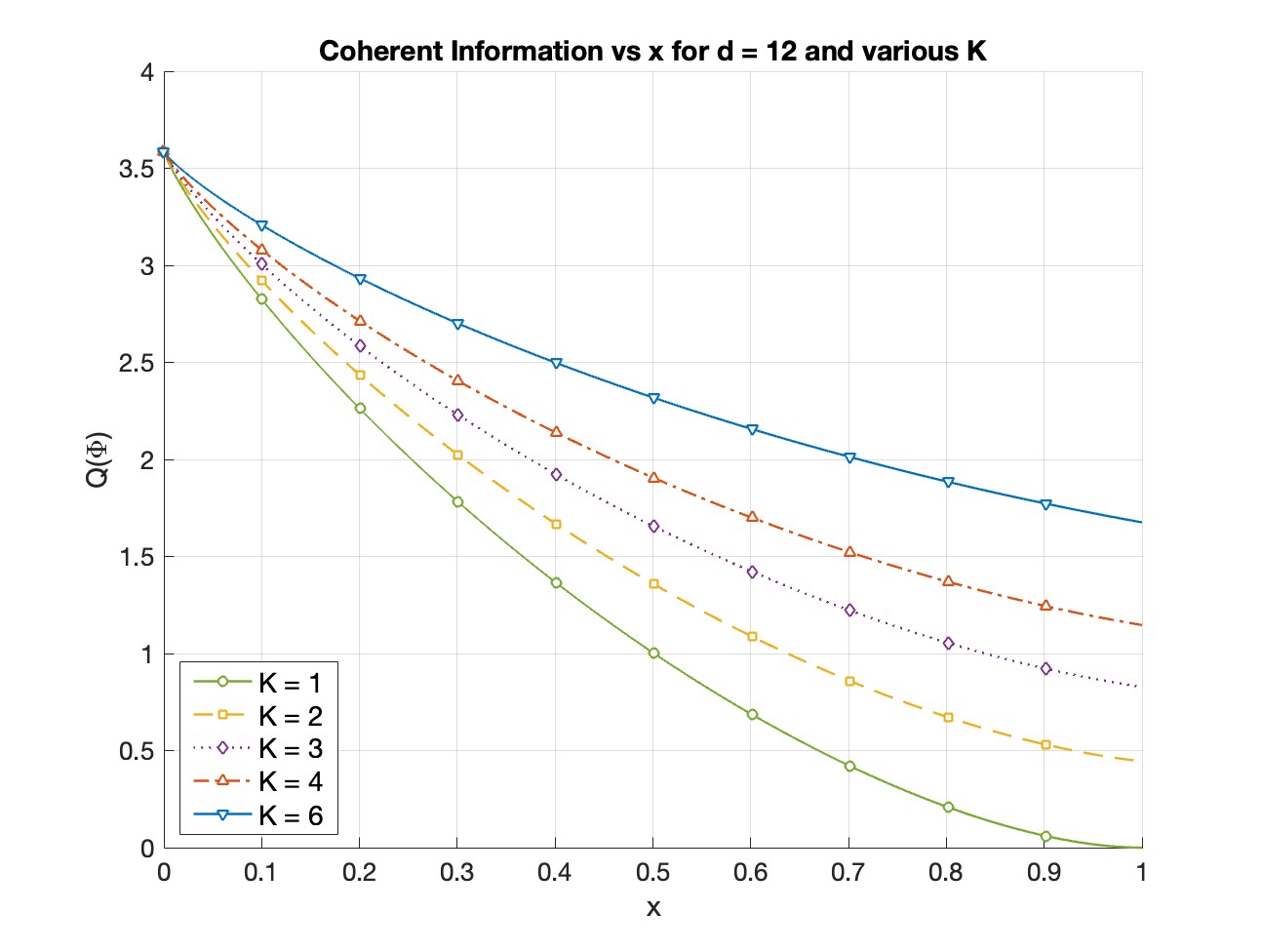}
	\caption{\tb{Coherent information $Q\bigl(\Phi\bigr)$ (eq. \ref{e40}) versus $x$ for $d=12$ and $k\in\{1,2,3,4,6\}$. It is seen that compared with figure (\ref{Capacities}),   here the capacity for each value of $k$, although still non-vanishing, is smaller than the correponding capacity of the block-decohering channel. This is due to the absence of any decoherence-free subspace in this case.}} 
	\label{fig:coherent-info}
\end{figure}
\ni As special cases we note that 
when $k=1$, this capacity reduces to that of (\ref{decorenceQcapa}), as expected.
 Also for $x=0$, or when $k=d$, for which  the channel is the identity channel, the capacity is given by $Q(\Phi_k)=\log_2 d. $  Note that for the strongest level of noise $(x=1)$, the quantum capacity is no longer given by $\log_2 k$.  
The capacities of these channels for $d=12$ and for various degrees of decoherence, determined by $k$ are plotted in figure (\ref{fig:coherent-info}). It is worth nothing that even at full decoherence, the channel has non-zero capacity, despite having no decoherence-free subspace.

\section{Discussion}
\label{discussion}

In this work, we have derived an exact, closed-form expression for the quantum capacity of a wide class of decohering channels in arbitrary dimensions. \tr{ These channels, which interpolate between ideal transmission and complete decoherence, are  degradable for all noise parameters, allowing their capacity to be computed without regularization. Our results expand the set of channels for which the capacity is computable exactly and in a compact analytical form. Remarkably these expressions are valid at arbitrary dimension, allowing a comprehension of the communication rates of higher-dimensional channels which largely have not yet been explored.}\\

\ni Beyond the exact formulas, our findings reveal a deeper structural insight: the geometry of decoherence—specifically the structure of coherence-preserving subspaces—plays a decisive role in determining whether a channel can sustain quantum information transmission, see also \cite{chessa_partially_2021}. We show that even fully decohering channels may retain nonzero quantum capacity, provided that the decoherence acts in a way that respects overlap between certain subspaces. This challenges the conventional assumption that maximal decoherence necessarily eliminates all capacity.\\

\ni The channels we analyze may serve as analytically tractable models for noise processes in a variety of physical platforms, including quantum memories, spin chains, and systems with partial access to coherence-preserving degrees of freedom. 
\tr{By providing an exact solution for the quantum capacity of the decohering channels, our work  connects theoretical capacity limits to the physical decoherence processes in real devices. This  allows for the design of quantum encoding schemes to combat structured decohering noise, which may be relevant for  the development of practical quantum memories and networks for the NISQ era \cite{kimble2008,brown_quantum_2016,preskill_quantum_2018,gisin_quantum_2007}.}\\

\ni  Finally, our work opens up several new directions in the problem of capacity of quantum channels. The first one is to use these results for obtaining upper or lower bounds for more general models of noise in arbitrary dimensions. \tr{As examples, one can obtain upper bounds for the capacity of channels which are concatenated by these decohering channesl, or  lower bounds for channels which are less noisy than the channels studied here.  }
 \tr{The second avenue is based on extending our method for proving degradability to more general channels. While the relation $\Lambda^c\circ \Lambda=\Lambda^c$, has been proved for generalized dephasing channel in previous works \cite{khatri2023positivity}, our method of breaking this relation into two relations $D\circ \Lambda=D$ and $\Lambda^c\circ D=\Lambda^c$, seems to be like reducing of a second order equation (in terms of the unknown Kraus operators) into two first order relations. This method may facilitate the search for those channels in which the complementary channel degrades the channel itself.} \\ 

\noindent {{\bf Acknowledgement:}} We thank  Stefano Chessa for his careful reading of the manuscript and for helpful correspondence, including drawing our attention to some relevant works. V.K. Thanks Dariusz Chruscinski, Saverio Pascazio, Gniewko Sarbicki and Mario Ziman for their valuable comments, when this work was presented in the 56th Symposium in Mathematical Physics in Torun, Poland. This work was supported by  Iran National Science Foundation, under Grants No.4022322 and QST 4040203.

	\bibliography{refs}

\begin{thebibliography}{10}

\bibitem{NielsenChuang2000}
Michael~A. Nielsen and Isaac~L. Chuang.
\newblock {\em Quantum Computation and Quantum Information}.
\newblock Cambridge University Press, 2000.

\bibitem{Lloyd1997}
Seth Lloyd.
\newblock Capacity of the noisy quantum channel.
\newblock {\em Physical Review A}, 55(3):1613--1622, 1997.

\bibitem{Schumacher_1996}
Benjamin Schumacher and M.~A. Nielsen.
\newblock Quantum data processing and error correction.
\newblock {\em Physical Review A}, 54(4):2629–2635, October 1996.

\bibitem{Shor2004}
Peter~W. Shor.
\newblock Equivalence of additivity questions in quantum information theory.
\newblock {\em Communications in Mathematical Physics}, 246:453--472, 2004.

\bibitem{Hastings_2009}
M.~B. Hastings.
\newblock Superadditivity of communication capacity using entangled inputs.
\newblock {\em Nature Physics}, 5(4):255–257, March 2009.

\bibitem{SmithYard2008}
Graeme Smith and Jon Yard.
\newblock Quantum communication with zero-capacity channels.
\newblock {\em Science}, 321(5897):1812--1815, 2008.

\bibitem{gambetta2010randomized}
Jay~M Gambetta, A~D C{\'o}rcoles, Seth~T Merkel, John~A Smolin, Blake Ware,
  Blake~R Johnson, David Puzzuoli, Thomas Ohki, Jim Rozen, David~W Abraham,
  et~al.
\newblock Randomized benchmarking for quantum channels.
\newblock {\em arXiv preprint arXiv:1004.0329}, 2010.

\bibitem{da2010direct}
Marcus~P Da~Silva, Olivier Landon-Cardinal, and David Poulin.
\newblock Direct fidelity estimation of quantum channels.
\newblock {\em arXiv preprint arXiv:1004.2335}, 2010.

\bibitem{macchiavello2004experimental}
C~Macchiavello, G~Massimo Palma, and S~Virmani.
\newblock Experimental quantum capacity of a channel with a correlated noise
  environment.
\newblock {\em Physical Review A}, 69(3):030302, 2004.

\bibitem{durkin2007experimental}
Gabriel~A Durkin, Carlos~C L{\'o}pez, Jeffrey~H Shapiro, and Jonathan~P
  Dowling.
\newblock Experimental demonstration of quantum channel capacity improvement
  using classical correlations.
\newblock {\em Physical review letters}, 99(2):020401, 2007.

\bibitem{Bennett_1997}
Charles~H. Bennett, David~P. DiVincenzo, and John~A. Smolin.
\newblock Capacities of quantum erasure channels.
\newblock {\em Physical Review Letters}, 78(16):3217–3220, April 1997.

\bibitem{devetak_capacity_2005}
I.~Devetak and P.~W. Shor.
\newblock The {Capacity} of a {Quantum} {Channel} for {Simultaneous}
  {Transmission} of {Classical} and {Quantum} {Information}.
\newblock {\em Communications in Mathematical Physics}, 256(2):287--303, June
  2005.

\bibitem{chessa_quantum_2021}
Stefano Chessa and Vittorio Giovannetti.
\newblock Quantum capacity analysis of multi-level amplitude damping channels.
\newblock {\em Communications Physics}, 4(1):22, February 2021.

\bibitem{Chessa_2023}
Stefano Chessa and Vittorio Giovannetti.
\newblock Resonant multilevel amplitude damping channels.
\newblock {\em Quantum}, 7:902, January 2023.

\bibitem{Shirokov2006}
M.~E. Shirokov.
\newblock The holevo capacity of infinite dimensional channels and the
  additivity problem.
\newblock {\em Communications in Mathematical Physics}, 262(1):137--159, 2006.

\bibitem{KarimipourSciRep2025}
Shayan Roofeh and Vahid Karimipour.
\newblock Phase transition in the quantum capacity of quantum channels.
\newblock 2024.
\newblock arXiv:2408.05733 [quant-ph].

\bibitem{karimipour2024}
Vahid Karimipour.
\newblock Noisy landau--streater and werner--holevo channels in arbitrary
  dimensions.
\newblock {\em Physical Review A}, 110(2):022424, 2024.

\bibitem{Wolf2007}
Michael~M. Wolf and David P{\'e}rez-Garc{\'i}a.
\newblock Quantum capacities of channels with small environment.
\newblock {\em Physical Review A}, 75(1):012303, 2007.

\bibitem{SmithSmolin2007}
Graeme Smith and John~A. Smolin.
\newblock Additive extensions of a quantum channel.
\newblock {\em IEEE Transactions on Information Theory}, 53(12):4733--4741,
  2007.

\bibitem{Cubitt2008}
Toby~S. Cubitt, Mary~Beth Ruskai, and Graeme Smith.
\newblock The structure of degradable quantum channels.
\newblock {\em Journal of Mathematical Physics}, 49(10):102104, 2008.

\bibitem{Fukuda2009}
Motohisa Fukuda and Michael~M. Wolf.
\newblock Simplifying additivity problems using direct sum constructions.
\newblock {\em Journal of Mathematical Physics}, 50(8):083503, 2009.

\bibitem{FanizzaKianvashGiov2020}
Marco Fanizza, Farzad Kianvash, and Vittorio Giovannetti.
\newblock Quantum flags and new bounds on the quantum capacity of the
  depolarizing channel.
\newblock {\em Physical Review Letters}, 125(2):020503, 2020.

\bibitem{Roofeh_2024}
Shayan Roofeh and Vahid Karimipour.
\newblock Noisy werner-holevo channel and its properties.
\newblock {\em Physical Review A}, 109(5), May 2024.

\bibitem{RoofehKarimipour2024}
Shayan Roofeh and Vahid Karimipour.
\newblock Capacities of a two-parameter family of noisy werner–holevo
  channels.
\newblock {\em Quantum Information Processing}, 2025.
\newblock to appear.

\bibitem{baumgratz_quantifying_2014}
T.~Baumgratz, M.~Cramer, and M. B. Plenio.
\newblock Quantifying {Coherence}.
\newblock {\em Physical Review Letters}, 113(14):140401, September 2014.

\bibitem{zurek2003}
Wojciech~H. Zurek.
\newblock Decoherence, einselection, and the quantum origins of the
  classical.
\newblock {\em Reviews of Modern Physics}, 75(3):715--775, 2003.

\bibitem{schlosshauer2005}
Maximilian Schlosshauer.
\newblock Decoherence, the measurement problem, and interpretations of quantum
  mechanics.
\newblock {\em Reviews of Modern Physics}, 76(4):1267--1305, 2005.

\bibitem{Mani_2015}
Azam Mani and Vahid Karimipour.
\newblock Cohering and decohering power of quantum channels.
\newblock {\em Physical Review A}, 92(3), September 2015.

\bibitem{ManiRezazadehKarimipour2024}
Azam Mani, Fatemeh Rezazadeh, and Vahid Karimipour.
\newblock Quantum coherence between subspaces: State transformation, cohering
  power, $k$‑coherence and other properties.
\newblock {\em Physical Review A}, 109(1):012435, 2024.

\bibitem{PhysRevLett.84.2525}
Emanuel Knill, Raymond Laflamme, and Lorenza Viola.
\newblock Theory of quantum error correction for general noise.
\newblock {\em Phys. Rev. Lett.}, 84:2525--2528, Mar 2000.

\bibitem{PhysRevLett.82.2417}
Lorenza Viola, Emanuel Knill, and Seth Lloyd.
\newblock Dynamical decoupling of open quantum systems.
\newblock {\em Phys. Rev. Lett.}, 82:2417--2421, Mar 1999.

\bibitem{PhysRevLett.79.3306}
P.~Zanardi and M.~Rasetti.
\newblock Noiseless quantum codes.
\newblock {\em Phys. Rev. Lett.}, 79:3306--3309, Oct 1997.

\bibitem{Lidar_2003}
Daniel~A. Lidar and K.~Birgitta~Whaley.
\newblock {\em Decoherence-Free Subspaces and Subsystems}, page 83–120.
\newblock Springer Berlin Heidelberg, 2003.

\bibitem{Duan_1997}
Lu-Ming Duan and Guang-Can Guo.
\newblock Preserving coherence in quantum computation by pairing quantum bits.
\newblock {\em Physical Review Letters}, 79(10):1953–1956, September 1997.

\bibitem{aberg_quantifying_2006}
Johan Aberg.
\newblock Quantifying {Superposition}, December 2006.
\newblock arXiv:quant-ph/0612146.

\bibitem{bischof_resource_2019}
Felix Bischof, Hermann Kampermann, and Dagmar Bruß.
\newblock Resource {Theory} of {Coherence} {Based} on
  {Positive}-{Operator}-{Valued} {Measures}.
\newblock {\em Physical Review Letters}, 123(11):110402, September 2019.

\bibitem{bischof_quantifying_2021}
Felix Bischof, Hermann Kampermann, and Dagmar Bruß.
\newblock Quantifying coherence with respect to general quantum measurements.
\newblock {\em Physical Review A}, 103(3):032429, March 2021.

\bibitem{levi_quantitative_2014}
Federico Levi and Florian Mintert.
\newblock A quantitative theory of coherent delocalization.
\newblock {\em New Journal of Physics}, 16(3):033007, March 2014.

\bibitem{Ringbauer2018}
Michael Ringbauer, Thomas~R. Bromley, Marco Cianciaruso, Ludovico Lami,
  W.~Y.~Samuel Lau, Gerardo Adesso, Andrew~G. White, Alessandro Fedrizzi, and
  Marco Piani.
\newblock Certification and quantification of multilevel quantum coherence.
\newblock {\em Physical Review X}, 8(4):041007, oct 2018.

\bibitem{Johnston2018}
Nathaniel Johnston, Chi-Kwong Li, Sarah Plosker, Yiu-Tung Poon, and Bartosz
  Regula.
\newblock Evaluating the robustness of \textit{k}-coherence and
  \textit{k}-entanglement.
\newblock {\em Physical Review A}, 98(2):022328, aug 2018.

\bibitem{xu_coherence_2020}
Jianwei Xu, Lian-He Shao, and Shao-Ming Fei.
\newblock Coherence measures with respect to general quantum measurements.
\newblock {\em Physical Review A}, 102(1):012411, July 2020.

\bibitem{kais_review_2014}
Daniel~A. Lidar.
\newblock Review of {Decoherence}‐{Free} {Subspaces}, {Noiseless}
  {Subsystems}, and {Dynamical} {Decoupling}.
\newblock In Sabre Kais, editor, {\em Advances in {Chemical} {Physics}}, pages
  295--354. Wiley, 1 edition, February 2014.

\bibitem{chessa_partially_2021}
Stefano Chessa and Vittorio Giovannetti.
\newblock Partially {Coherent} {Direct} {Sum} {Channels}.
\newblock {\em Quantum}, 5:504, July 2021.

\bibitem{barnum_information_1998}
Howard Barnum, M.~A. Nielsen, and Benjamin Schumacher.
\newblock Information transmission through a noisy quantum channel.
\newblock {\em Physical Review A}, 57(6):4153--4175, June 1998.

\bibitem{lloyd_capacity_1997}
Seth Lloyd.
\newblock Capacity of the noisy quantum channel.
\newblock {\em Physical Review A}, 55(3):1613--1622, March 1997.

\bibitem{devetak_private_2005}
I.~Devetak.
\newblock The private classical capacity and quantum capacity of a quantum
  channel.
\newblock {\em IEEE Transactions on Information Theory}, 51(1):44--55, January
  2005.

\bibitem{cubitt_2015}
Toby Cubitt, David Elkouss, William Matthews, Maris Ozols, David
  Pérez-García, and Sergii Strelchuk.
\newblock Unbounded number of channel uses may be required to detect quantum
  capacity.
\newblock {\em Nature Communications}, 6(1):6739, March 2015.

\bibitem{fern_lower_2008}
Jesse Fern and K.~Birgitta Whaley.
\newblock Lower bounds on the nonzero capacity of {Pauli} channels.
\newblock {\em Physical Review A}, 78(6):062335, December 2008.

\bibitem{divincenzo_quantum-channel_1998}
David~P. DiVincenzo, Peter~W. Shor, and John~A. Smolin.
\newblock Quantum-channel capacity of very noisy channels.
\newblock {\em Physical Review A}, 57(2):830--839, February 1998.

\bibitem{leditzky_quantum_2018}
Felix Leditzky, Debbie Leung, and Graeme Smith.
\newblock Quantum and {Private} {Capacities} of {Low}-{Noise} {Channels}.
\newblock {\em Physical Review Letters}, 120(16):160503, April 2018.

\bibitem{bausch_error_2021}
Johannes Bausch and Felix Leditzky.
\newblock Error {Thresholds} for {Arbitrary} {Pauli} {Noise}.
\newblock {\em SIAM Journal on Computing}, 50(4):1410--1460, January 2021.

\bibitem{siddhu_positivity_2021}
Vikesh Siddhu and Robert~B. Griffiths.
\newblock Positivity and {Nonadditivity} of {Quantum} {Capacities} {Using}
  {Generalized} {Erasure} {Channels}.
\newblock {\em IEEE Transactions on Information Theory}, 67(7):4533--4545, July
  2021.

\bibitem{smith_additive_2007}
Graeme Smith and John~A. Smolin.
\newblock Additive {Extensions} of a {Quantum} {Channel}, December 2007.
\newblock arXiv:0712.2471 [quant-ph].

\bibitem{bhalerao2023improving}
Sujeet Bhalerao and Felix Leditzky.
\newblock Improving quantum communication rates with permutation-invariant
  codes.
\newblock In {\em 2023 IEEE International Symposium on Information Theory
  (ISIT)}, pages 108--112. IEEE, 2023.

\bibitem{kianvash1}
Marco Fanizza, Farzad Kianvash, and Vittorio Giovannetti.
\newblock Quantum {Flags} and {New} {Bounds} on the {Quantum} {Capacity} of the
  {Depolarizing} {Channel}.
\newblock {\em Physical Review Letters}, 125(2):020503, July 2020.

\bibitem{kianvash2}
Marco Fanizza, Farzad Kianvash, and Vittorio Giovannetti.
\newblock Estimating {Quantum} and {Private} capacities of {Gaussian} channels
  via degradable extensions.
\newblock {\em Physical Review Letters}, 127(21):210501, November 2021.
\newblock arXiv:2103.09569 [quant-ph].

\bibitem{poshtvan_capacities_2022}
Abbas Poshtvan and Vahid Karimipour.
\newblock Capacities of the covariant {Pauli} channel.
\newblock {\em Physical Review A}, 106(6):062408, December 2022.

\bibitem{Chessa_2019}
Stefano Chessa, Marco Fanizza, and Vittorio Giovannetti.
\newblock Quantum-capacity bounds in spin-network communication channels.
\newblock {\em Physical Review A}, 100(3), September 2019.

\bibitem{cubitt_structure_2008}
Toby~S. Cubitt, Mary~Beth Ruskai, and Graeme Smith.
\newblock The structure of degradable quantum channels.
\newblock {\em Journal of Mathematical Physics}, 49(10):102104, October 2008.

\bibitem{khatri2023positivity}
S.~Khatri, G.~Lami, and M.~M. Wilde.
\newblock Positivity of quasi-relative entropy under commutative conditional
  expectations.
\newblock {\em arXiv preprint arXiv:2309.13758}, 2023.

\bibitem{smaczynski2016selfcomplementary}
Marek Smaczyński, Wojciech Roga, and Karol Życzkowski.
\newblock Selfcomplementary {Quantum} {Channels}.
\newblock {\em Open Systems \& Information Dynamics}, 23(03):1650014, September
  2016.

\bibitem{Schur1923}
Issai Schur.
\newblock {\"U}ber eine {Klasse} von {Mittelbildungen} mit {Anwendungen} auf
  die {Determinantentheorie}.
\newblock {\em Sitzungsberichte der Berliner Mathematischen Gesellschaft},
  22:9--20, 1923.

\bibitem{Horn1954}
Alfred Horn.
\newblock Doubly stochastic matrices and the diagonal of a rotation matrix.
\newblock {\em American Journal of Mathematics}, 76:620--630, 1954.

\bibitem{holevo2005complementarychannelsadditivityproblem}
A.~S. Holevo.
\newblock On complementary channels and the additivity problem, 2005.

\bibitem{Sperling2015}
J.~Sperling and W.~Vogel.
\newblock Convex ordering and quantification of quantumness.
\newblock {\em Physica Scripta}, 90(7):074024, jul 2015.

\bibitem{Killoran2016}
Nathan Killoran, Frank E.~S. Steinhoff, and Martin~B. Plenio.
\newblock Converting nonclassicality into entanglement.
\newblock {\em Physical Review Letters}, 116(8):080402, feb 2016.

\bibitem{johnston_evaluating_2018}
Nathaniel Johnston, Chi-Kwong Li, Sarah Plosker, Yiu-Tung Poon, and Bartosz
  Regula.
\newblock Evaluating the robustness of k -coherence and k -entanglement.
\newblock {\em Physical Review A}, 98(2):022328, August 2018.

\bibitem{senitzky_incoherence_1962}
I.~R. Senitzky.
\newblock Incoherence, {Quantum} {Fluctuations}, and {Noise}.
\newblock {\em Physical Review}, 128(6):2864--2870, December 1962.

\bibitem{kimble2008}
H.~Jeff Kimble.
\newblock The quantum internet.
\newblock {\em Nature}, 453:1023--1030, 2008.

\bibitem{brown_quantum_2016}
Benjamin~J. Brown, Daniel Loss, Jiannis~K. Pachos, Chris~N. Self, and James~R.
  Wootton.
\newblock Quantum memories at finite temperature, November 2016.
\newblock arXiv:1411.6643.

\bibitem{preskill_quantum_2018}
John Preskill.
\newblock Quantum {Computing} in the {NISQ} era and beyond, July 2018.
\newblock arXiv:1801.00862 version: 3.

\bibitem{gisin_quantum_2007}
Nicolas Gisin and Rob Thew.
\newblock Quantum communication.
\newblock {\em Nature Photonics}, 1(3):165--171, March 2007.

\end{thebibliography}
	\pagebreak

\end{document}